\documentstyle[12pt]{article}
\renewcommand{\baselinestretch}{1.4}

\begin{document}

\begin{flushright}
IPNO/TH 95-57
\end{flushright}
\vspace{1 cm}
\begin{center}
{\large {\bf The two-fermion vector potential of constraint theory \\
from Feynman diagrams}}
\vspace{1 cm}

{H. Jallouli and H. Sazdjian}\\
{\it Division de Physique Th\'eorique\footnote{Unit\'e de Recherche
des Universit\'es Paris 11 et Paris 6 associ\'ee au CNRS.},
Institut de Physique Nucl\'eaire,\\
Universit\'e Paris XI,\\
F-91406 Orsay Cedex, France}
\vspace{3 cm}

{\large Abstract}
\end{center}
The relativistic fermion-antifermion bound state vector potential of
constraint theory is calculated, in perturbation theory,
by means of the Lippmann$-$Schwinger
type equation that relates it to the scattering amplitude.
Leading contributions of $n$-photon exchange diagrams are calculated
in an approximation scheme that adapts eikonal approximation to the
bound state problem. They produce terms proportional, in
three-dimensional $x$-space, to $(\alpha /r)^n$.
The series of leading contributions is summed.

\newpage

We report in this note results on calculations, in the framework of
the manifestly covariant formalism of
constraint theory \cite{ll,cva1,s1}, of the relativistic two-body
bound state vector potential from summation of Feynman diagrams,
considered in their leading order of the infra-red counting rules
of QED.\par
Perturbative calculations of the three-dimensional potential from the
Bethe$-$Salpeter kernel or from the scattering amplitude have been
successful in the past mainly in the Coulomb gauge for the exchanged
photon propagator \cite{by}. Covariant gauges \cite{l}, or
covariant propagators for scalar exchanges \cite{bcm}, produce
spurious infra-red singularities that are only cancelled by
contributions of higher order diagrams. This defect seems to be
related to the noncovariant nature of the three-dimensional equation
around which iteration of the Bethe$-$Salpeter equation is accomplished.
It turns out, as we shall see below, that in the manifestly covariant
formalism of constraint theory these complications are absent: in each
formal order of perturbation theory the leading term of the potential
due to ladder and crossed ladder diagrams
is free of spurious singularities. A similar phenomenon also occurs in
the variant of the quasipotential approach developed by Todorov \cite{t}.
In particular, it has been shown \cite{rta}, for one spin-1/2 and one
spin-0 particle systems, that in the two-photon exchange diagrams,
calculated in the Feynman gauge, the spurious infra-red singularities
cancel out.\par
We concentrate in the following on the case of bound states composed
of one fermion with mass $m_1$ and one antifermion with mass $m_2$,
mutually interacting by means of vector (massless) photons.
\par
We use standard notations for the total and relative variables:
$P = p_1+p_2$, $p = (p_1 -p_2)/2$, $x = x_1 - x_2$.
For states that are eigenstates of the total momentum $P$ we define
transverse and longitudinal decompositions of four-vectors with respect
to $P$ and denote them with indices $T$ and $L$, respectively.
A detailed presentation of the constraint theory wave equations, of
their properties, as well as of the way they can be reduced to a single
Pauli$-$Schr\"{o}dinger type equation can be found in Ref. \cite{ms}.
The wave equations determine, among others, the c.m. relative time
evolution of the system and allow the elimination of the c.m. relative
energy. These properties are consequences of the following constraint
equation:
\begin{equation} \label{e3}
C(p)\ \equiv
(p_1^2 - p_2^2)-(m_1^2 - m_2^2)\ =\ 2P_L p_L - (m_1^2 - m_2^2)
\ \approx \ 0\ .
\end{equation}
When constraint $C$ is used, the individual
Klein$-$Gordon operators become equal:
\begin{equation} \label{e5}
H_0\ \equiv \ (p_1^2 - m_1^2)\bigg \vert _C \ =\ (p_2^2 - m_2^2)
\bigg \vert _C \ =\ \frac {P^2}{4} - \frac {1}{2}(m_1^2 + m_2^2)
+ \frac {(m_1^2 - m_2^2)^2}{4P^2} + p^{T2}\ .
\end{equation}
\par
The potential $\widetilde V$ that appears in the wave equations must
be compatible with constraint $C$ [Eq. (\ref{e3})]. This implies that
it should not depend on the relative longitudinal coordinate $x_L$:
$\widetilde V \ =\ \widetilde V (x^T, P_L, p^T, \gamma _1, \gamma _2)$ .
[$\gamma _1$ and $\gamma _2$ are the Dirac matrices for particles 1 and
2, respectively. In general, $\widetilde V$ is an integral operator in
$x^T$.] Thus, the internal dynamics of the system is three-dimensional,
apart from the spin degrees of freedom.\par
The wave equations of constraint theory can be connected to the
Bethe$-$Salpeter equation, by iterating the latter around the constraint
hypersurface (\ref{e3}) \cite{s2}. This connection determines the
relationship of the potential $\widetilde V$ with the off-mass shell
scattering amplitude, through a Lippmann$-$Schwinger type equation:
\begin{equation} \label{e7}
\widetilde V\ =\ \widetilde T + \widetilde V \widetilde g_0
\widetilde T\ ,\ \ \ \ \
\widetilde T = \frac {i}{2P_L}\bigg [ T(P,p,p') \bigg ]_
{C(p),C(p')}\ ,
\end{equation}
where $T$ is the off-mass shell scattering amplitude in which
the external momenta are submitted to the constraint (\ref{e3});
$\widetilde g_0$ is defined as:
\begin{equation} \label{e9}
\widetilde g_0\ =\ \bigg [ S_1(p_1) S_2(-p_2) H_0 \bigg ]_{C(p)}\ ,
\end{equation}
where $S_1$ and $S_2$ are the propagators of the fermions (submitted
in Eq. (\ref{e9}) to the constraint (\ref{e3})) and $H_0$ is defined
in Eq. (\ref{e5}).
\par
Equation (\ref{e7}) is the basis for the calculation of the potential in
terms of Feynman diagrams. We limit ourselves to the evaluation of the
ladder and crossed ladder diagrams (in their leading order), neglecting
the contributions of radiative corrections.
Iteration of Eq. (\ref{e7}) yields for the potential the expansion:
\begin{equation} \label{e11}
\widetilde V \equiv\ \sum_{n=1}^{\infty} \widetilde V^{(n)}\ =\
\widetilde T \sum _{p=0}^{\infty} (\widetilde g_0 \widetilde T)^p\ ,
\end{equation}
where $\widetilde V^{(n)}$ is that part of the potential that results
from $n$ exchanged photons. We observe that the perturbation series
of $\widetilde V$ contains, in addition to the usual Feynman
diagrams, other types of diagram arising from the presence of the
constraint factor $\widetilde g_0$; we shall call these diagrams
``constraint diagrams''; they are obtained from the usual box-ladder
type diagrams by the replacement of (at least) one pair of fermion
and antifermion propagators by the constraint factor $2i\pi \delta
(C) \widetilde g_0$. The role of these diagrams will be to cancel
the spurious infra-red singularities coming from the amplitude.
The diagrams corresponding to two photon exchanges are represented
in Fig. 1. [The constraint diagram is denoted by a cross.]\par
When constraint (\ref{e3}) is imposed on the external particles, one
obtains: $q_L=p_{1L}-p_{1L}'=0$, $q^2=(p_1-p_1')^2=q^{T2}$.
The Feynman diagrams are calculated with external particles considerd off
the mass shell, with their longitudinal momenta fixed by the bound state
mass $P_L$ [the binding energy is of order $\alpha ^2$], through
Eq. (\ref{e3}), while the transverse momenta $p^T$ and $p'^T$ have,
according to the infra-red counting rules of QED, orders of
magnitude of $\alpha $. The photon is taken massless.\par
The potential $\widetilde V^{(1)}$ that results from one
photon exchange is in the Feynman gauge, in three-dimensional $x$-space:
\begin{equation} \label{e21}
\widetilde V^{(1)}\ =\ \frac {\alpha }{2P_L r} \gamma _1.\gamma _2
\ ,\ \ \ \alpha \ =\ \frac {e^2}{4\pi }\ ,\ \ \ r\ =\
\sqrt {-x^{T2}}\ .
\end{equation}
[The  matrices $\gamma _2$,
corresponding to the antifermion, act on the $4\times 4$ matrix wave
function from the right.]\par
The vector potential can be divided into two parts: the timelike
component, proportional to $\gamma _{1L} \gamma _{2L}$, and the
spacelike component, proportional to $\gamma _1^{T\mu }\gamma _2^{T\nu}$.
Because of the latter matrices, which are of order $\alpha ^2$, the
spacelike component, at each formal order of perturbation theory,
will be damped with respect to the timelike component by an
$O(\alpha ^2)$ factor. In the present work, where we are calculating
only leading terms of higher order diagrams, we can evaluate only the
timelike component of the vector potential. However, if one assumes
that in the Feynman gauge a simple structure of the $\gamma$-matrices
like that of Eq. (\ref{e21}) survives
in higher orders, then the knowledge of the timelike component will
still allow one to reconstitute the whole potential in this gauge.
\par
The potential $\widetilde V^{(2)}$ is calculated
by evaluating the diagrams of Fig. 1. The details of the calculations
will be presented
elsewhere; we also refer the reader to Ref. \cite{rta}, where
analogous calculations are done and similar properties to those obtained
here found. The various integrals can be calculated to the desired
precision by dividing the integration domains into several intervals,
in which appropriate approximations can be used.\par
To understand the various cancellation mechanisms, we explicitly write
the product of the two fermion propagators occurring in the box
diagram of Fig. 1, where $k$ is the four-momentum carried by one of
the photons:
\begin{eqnarray} \label{e22}
S_1(p_1-k) S_2(-(p_2+k))\ =\ \frac {i}{(p_1-k)^2 - m_1^2 +i\epsilon}\
\frac {i}{(p_2+k)^2 - m_2^2 +i\epsilon} \times\nonumber \\
\bigg [ (\gamma _1.p_1 + m_1) (-\gamma _2.p_2 + m_2)
-(\gamma _1.p_1 + m_1) \gamma _2.k - (-\gamma _2.p_2 + m_2)
\gamma _1.k + \gamma _1.k \gamma _2.k \bigg ]\ . \nonumber \\
\end{eqnarray}
\par
The first term in the brackets yields a contribution to the potential
(in $x$-space) composed of two terms, of orders $\alpha ^2 \ln \alpha ^{-1}$
and $\alpha ^3 \ln \alpha ^{-1}$, respectively. The analogous contribution
of the crossed diagram is of order $\alpha ^3 \ln \alpha ^{-1}$. The
corresponding contribution of the constraint diagram is of order
$\alpha ^2 \ln \alpha ^{-1}$ and cancels the similar term of the box
diagram. The remaining term of the box diagram cancels the
$O(\alpha ^3 \ln \alpha ^{-1})$ term of the crossed diagram and yields
a term of order $\alpha ^3$, which in turn can be removed by a finite
multiplicative renormalization of the constraint factor (\ref{e9})
[we use the prescription that the $1/r$-terms of the potentials should
arise only from the one-photon exchange diagram]. The sum of the
above terms is of order $\alpha ^5 \ln \alpha ^{-1}$, which, compared
to the leading $O(\alpha ^4)$ effect we are estimating, can be neglected.
\par
The second and third terms in the brackets of Eq. (\ref{e22}) and the
similar terms of the crossed and constraint diagrams yield
$O(\alpha ^4 \ln \alpha ^{-1})$ and $O(\alpha ^4)$ terms. The
$O(\alpha ^4 \ln \alpha ^{-1})$ terms are mutually cancelled and one
remains with the $O(\alpha ^4)$ terms. The latter actually arise
from the longitudinal component $k_L$ contributions of the photon
momentum $k$ multiplying the $\gamma $-matrices.\par
The last term in the brackets in Eq. (\ref{e22}) and the similar ones
of the other diagrams are individually negligible in front of the
$O(\alpha ^4)$ terms.\par
For the evaluation of the $O(\alpha ^4)$ terms, one can replace at
leading order the matrices $\gamma _{1L}$ and $\gamma _{2L}$ by their
eigenvalues, $+1$ and $-1$, respectively, with respect to the
dominant component of the wave function, as well as $m_1$ by $p_{1L}$
and $m_2$ by $p_{2L}$. One finds:
\begin{equation} \label{e26}
\widetilde V^{(2)}\ =\ -
\frac {\alpha ^2}{2 P_L^2 r^2}\gamma _{1L} \gamma _{2L}\ .
\end{equation}
\par
It can be shown \cite{ms} that, when this contribution is incorporated
in the wave equations, the latter yield to order $1/c^2$,
up to wave function transformations, the Breit Hamiltonian,
which is known to produce the correct $O(\alpha ^4)$ effects. This
means that no other $O(\alpha ^4)$ terms should arise from higher
order diagrams.\par
It is natural to expect that the cancellation mechanism that occurs
at the level of two-photon exchange diagrams and yields at leading order
a local potential generalizes to higher orders. It does not seem possible
to verify this property in a rigorous way, due to the complexity of the
structure of $n$-photon diagrams. However, within a plausible
approximation scheme, that takes into account the properties of the
bound state we are considering, it is possible to show the above
property.\par
To device the appropriate approximation scheme, we make the following
observations: i) In result (\ref{e26}) the factor $1/r^2$
represents in momentum space the three-dimensional convolution (up to
multiplicative coefficients) of the two photon propagators, in
which the longitudinal component $k_L^2$ has been neglected; this
means that the entire $q^2$ dependence of the potential (at leading
order) comes from the photons and not from the fermions. ii)
$\widetilde V^{(2)}$ is independent of the off-mass shell condition
imposed on the fermions (i.e., does not depend on $(p_1^2 - m_1^2)$
or $(p_2^2 - m_2^2)$). iii) In the product of fermion propagators
[Eq. (\ref{e22})], only the linear longitudinal component terms $k_L$ of
the numerator have contributed to $\widetilde V^{(2)}$. These
observations suggest the following approximation of the fermion
propagators:
\begin{eqnarray} \label{e27}
S_1(p_1-k_1) &\simeq & \frac {i}{-2p_1.k_1 + i\epsilon }\
\big [ (\gamma _{1L} p_{1L} + m_1) - \gamma _{1L} k_{1L} \big ]\ ,\nonumber \\
S_2(-(p_2+k_2)) &\simeq & \frac {i}{2p_2.k_2 + i\epsilon }\
\big [ (-\gamma _{2L}p_{2L} + m_2) - \gamma _{2L} k_{2L} \big ]\ .
\end{eqnarray}
In the numerators, we have neglected the transverse momenta $p^T$
and $k_i^T$ (but {\it not} in the denominators); in
products of two fermion propagators we neglect in the numerator
quadratic terms $k_L^2$ (of the same $k_L$); we also neglect the total
momentum transfer $q$. In photon
propagators, we neglect $k_L^2$ dependences when they appear in
nonleading terms. Approximation (\ref{e27})
may be considered as a variant of the eikonal approximation
\cite{cw,ls}, adapted to the bound state problem.
Finally, because of the on-mass shell treatment of the external fermions
in approximation (\ref{e27}), the photon should be given a small
mass to prevent infra-red divergences at intermediate stages.\par
It can be checked by direct calculation that
the above approximations globally produce the correct
result (\ref{e26}).
\par
We now apply approximation (\ref{e27}) to $n$-photon exchange diagrams
$(n\geq 2)$. The main formula we use is a generalization of an identity
already used in the eikonal approximation (cf. Ref. \cite{ls}, Appendix).
Let $(c_1,c_2,\cdots ,c_n)$ be a set of $n$ numbers; it can be divided
into $n$ independent subsets of $(n-1)$ numbers, where in each subset one
of the $c_i$'s ($i=1,\cdots ,n)$ is missing. We have the identity:
\begin{equation} \label{e28}
\sum _{i=1}^n \sum _{perm} \bigg [ (c_1')^{-1} (c_1'+c_2')^{-1} \cdots
(c_1'+c_2'+\cdots +c_{n-1}')^{-1} \bigg ]_{c_i}\ =\
\left ( \sum _{i=1}^n c_i\right ) \bigg /
\left ( \prod _{j=1}^n c_j\right )\ ,
\end{equation}
where $(c_1',c_2',\cdots ,c_{n-1}')$ is a permutation of the subset of
$(n-1)$ numbers $(c_1,c_2,\cdots ,c_{i-1},$ $c_{i+1},\cdots ,c_n)$.
\par
Let us consider an $n$-photon exchange diagram (Fig. 2).
We denote by $k_1,\cdots ,k_n$ the momenta carried
by the photons. To take into account momentum conservation, a factor
$(2\pi )^4 \delta ^4(\sum _{i=1}^n k_i - q)$ must be included in the
corresponding integral. Since $q_L=0$, we have $\sum_{i=1}^n k_{iL}
=0$; but according to the approximations (\ref{e27}) we may also
neglect $q^T$ in the fermion propagators; hence, we have there the
approximation
\begin{equation} \label{e29}
\sum_{i=1}^n k_i\ =\ 0\ .
\end{equation}
The total number of the diagrams considered above is $n!$. In a given
diagram we have $(n-1)$ propagators of fermion 1 and $(n-1)$
propagators of fermion 2; only $(n-1)$ photon momenta appear on each of the
fermion
lines (but not the same in general). We first consider the terms that
do not contain any $k_{iL}$ $(i=1,\cdots ,n)$ in the numerator
[Eq. (\ref{e27})]. The corresponding $n!$ diagrams
can be divided into $n$ sets, where in each set we have $(n-1)!$
permutations of fermion 2 propagators (without the $k_{iL}$ factors
in the numerator) containing the same set of $(n-1)$ photon momenta.
With approximations (\ref{e27}), the sum of all these
propagators has the structure of the left-hand side of Eq. (\ref{e28}),
with $c_i = (2p_2.k_i+i\epsilon)$ [the terms $i(-\gamma _{2L} p_{2L}
+ m_2)$ are factorized]; hence, we obtain the factor
\begin{equation} \label{e30}
I_{2,n}\ =\ \frac {{\displaystyle \sum _{i=1}^n (2p_2.k_i+i\epsilon )}}
{{\displaystyle \prod _{j=1}^n (2p_2.k_j+i\epsilon )}}
\ =\ (-2i\pi)^{n-1} \delta (2p_2.k_1) \delta (2p_2.(k_1+k_2))\cdots
\delta (2p_2.\sum _{i=1}^{n-1} k_i)\ ,
\end{equation}
the second equality resulting from the use of Eq. (\ref{e29}) and
several algebraic operations (for more details cf. Ref. \cite{b},
pp. 116-117 and also Ref. \cite{cw}). The $\delta $-functions can
then be used, upon
integrations on the $k_{iL}$'s, in the fermion 1 propagators to
yield a global factor $(2P_L)^{-(n-1)} (\gamma _{1L} p_{1L} + m_1)
^{n-1} (-\gamma _{2L} p_{2L} + m_2)^{n-1}$. Denoting by $J^{(n)}$
the total contribution of the above diagrams
(without the coupling constants and other multiplicative coefficients),
we obtain:
\begin{eqnarray} \label{e31}
J^{(n)} &=& \left (\frac {-i}{2P_L}\right )^{n-1}\
\big [ (\gamma _{1L} p_{1L} +
m_1) (-\gamma _{2L} p_{2L} + m_2) \big ]^{n-1}\
(\gamma _{1L} \gamma _{2L})^n \nonumber \\
& & \times \int \big [ \prod _{i=1}^n \frac {d^3k^T}{(2\pi )^3}
\frac {1}{(k_i^{T2} - \mu ^2 + i\epsilon)} \big ]\ (2\pi )^3
\delta ^3(\sum _{i=1}^n k_i^T - q^T) \nonumber \\
& & \times \frac {1}{(-2p^T.k_1^T + i\epsilon )}\
\frac {1}{(-2p^T.(k_1^T+k_2^T) + i\epsilon )}\cdots
\frac {1}{(-2p^T.\sum _{j=1}^{n-1} k_j^T + i\epsilon )}\ ,
\nonumber \\
& &
\end{eqnarray}
where $\mu $ is a small mass given to the photon, to prevent infra-red
divergence. Also, the $\delta $-functions (\ref{e30}) yield for the
$k_{iL}^2$-terms orders of magnitude of $O(\alpha ^4)$, which accounts
for their omission in front of $O(\alpha ^2)$-terms in
approximations (\ref{e27}) and in photon propagators.
\par
On the other hand, we should also take into account the contributions of
constraint diagrams that are associated with the diagrams considered
above. A typical diagram where the constraint factor $\widetilde g_0$
[Eq. (\ref{e11})] appears twice is shown in Fig. 3.
The analysis of these diagrams is very similar to that already done in
the infinite mass limit (cf. Ref. \cite{ms}, Appendix C).
Each constraint diagram contributes the same quantity $J^{(n)}$ above
up to a sign factor equal to $(-1)^p$ if $\widetilde g_0$ occurs $p$
times in expansion (\ref{e11}). Taking into account the combinatorial
factors one finds that the sum of constraint diagrams cancels the
contribution $J^{(n)}$ of ordinary diagrams.
\par
This result can be generalized to the contributions of those parts of
the above diagrams where some of the $k_{iL}$'s (but not all) appear
in the numerator. These parts have also constraint diagram
counterparts and are cancelled by them.\par
Therefore, the only surviving parts of the above diagrams are those
where products of $(n-1)$ independent combinations of the $k_{iL}$'s
appear in the numerator; these parts do not have constraint diagram
counterparts, since there $k_{iL}=0$ $(i=1,\cdots ,n)$. After the
appearance of the $\delta $-functions from the permutational procedure
[Eq. (\ref{e30})], realized, according to the origin of the $k_{iL}$'s,
partly on line 1 and partly on line 2, the $k_{iL}$'s cancel by
integration, up to multiplicative coefficients, the remaining
fermion propagators, and one obtains a three-dimensional convolution
of $n$ photon propagators, yielding in $x$-space a potential
proportional to $(\alpha /r)^n$.
The analysis is much simplified by using these cancellations
(with the appropriate coefficients) prior to the permutational
procedure.
\par
We next turn to the calculation of the combinatorial factor associated
with the above procedure. In the product of the $(n-1)$ pairs of
fermion and antifermion propagators of a given diagram we can select
$p$ $k_{iL}$'s on line 1 and the complementary $(n-1-p)$ $k_{iL}$'s
on line 2. The choice of $p$ $k_{iL}$'s on line 1 can be made in
$\left (_{{\displaystyle \ \ \ p\ }}^{{\displaystyle n-1}}\right )$
different ways; however no freedom
is left for the choice of the set of $(n-1-p)$ complementary
$k_{iL}$'s on line 2.
During the permutational procedure, we need $(p+1)!$ permutations to
obtain $p$ $\delta $-functions on line 2 from the propagators not
having $k_{iL}$'s in their numerator; similarly, we need $(n-p)!$
permutations to produce $(n-1-p)$ $\delta $-functions on line 1.
After the approximations $(\gamma _{1L} p_{1L}+m_1)\simeq 2p_{1L}$
and $(-\gamma _{2L} p_{2L}+m_2)\simeq 2p_{2L}$ are made, the
integrations on the $k_{iL}$'s produce the global factor
$((-2i\pi)/(2P_L)) ^{n-1}$ .\par
Taking into account the total number of the diagrams with $n$
exchanged photons, which is $n!$, we obtain for the combinatorial
factor:
\begin{equation} \label{e34}
C_n\ =\ \sum _{p=0}^{n-1}
\left (_{{\displaystyle \ \ \ p\ }}^{{\displaystyle n-1}}\right )
\frac {n!} {(p+1)! (n-p)!}\ =\ F\left (-(n-1),-n;2;1\right )\ =\
\frac {(2n)!}{(n+1)! n!}\ ,
\end{equation}
where $F(a,b;c;z)$ is the hypergeometric function \cite{be}, the
value of which for $z=1$ is known in terms of $\Gamma $-functions.\par
The corresponding potential then becomes:
\begin{eqnarray} \label{e35}
\widetilde V^{(n)} &=& \frac {i}{2P_L} (-ie)^{2n} i^{2(n-1)}
(-i)^n \left ( \frac {-i}{2P_L}\right )^{n-1} \frac {(2n)!}{(n+1)! n!}
\ (\gamma _{1L} \gamma _{2L})^n \nonumber \\
& &\nonumber \\
& &\times \int \big [ \prod _{i=1}^n \frac {d^3k_i^T}{(2\pi)^3}
\frac {1}{(k_i^{T2}+i\epsilon)} \big ]\ (2\pi)^3 \delta ^3
(\sum _{j=1}^n k_j^T - q^T)\ ,
\end{eqnarray}
yielding in $x$-space:
\begin{eqnarray} \label{e36}
\widetilde V^{(n)}\ =\ -\frac {(2n)!}{(n+1)! n!} \left (-
\frac {\alpha }{2P_L r} \right )^n \gamma _{1L} \gamma _{2L}\ .
\\ \nonumber
\end{eqnarray}
[We have replaced $(\gamma _{1L} \gamma _{2L})^{n-1}$ by $(-1)^{n-1}$.]
\par
This expression is also valid for the one-photon exchange case
[Eq. (\ref{e21})], for which it provides the exact result,
concerning the timelike component.\par
The total timelike potential becomes:
\begin{equation} \label{e37}
\widetilde V\ =\ -\gamma _{1L} \gamma _{2L} \sum _{n=1}^{\infty }
(-1)^n \frac {(2n)!}{(n+1)! n!} \left ( \frac {\alpha }{2P_L r}\right )^n
\ =\  -\gamma _{1L} \gamma _{2L}
\left (\frac {1 - \sqrt {1+\frac {2\alpha }{P_L r}}}
{1 + \sqrt {1+\frac {2\alpha }{P_L r}}}\right )\ .
\end{equation}
\par
A parametrization that allows external field
interpretation of the potentials and ensures positivity of the norm
for energy independent potentials was proposed by Crater and
Van Alstine \cite{cva1} and used in Ref. \cite{ms}; it is:
$ \widetilde V\ =\ \tanh V $. Potential $\widetilde V$ above fits
this structure. Writing
\begin{equation} \label{e41}
\widetilde V\ =\ \tanh (\gamma _{1L} \gamma _{2L} V_2)\ ,
\end{equation}
we identify
\begin{equation} \label{e43}
V_2\ =\ \frac {1}{4} \ln \left (1+\frac {2\alpha }{P_L r}
\right )\ .
\end{equation}
\par
Furthermore, if we adopt, for the vector potential in the
Feynman gauge, the structure (\ref{e21}) of the $\gamma $-matrices,
obtained in lowest order, by the substitution $\gamma _{1L}
\gamma _{2L}$$\rightarrow $$\gamma _1.\gamma _2$, we can
reconstitute the whole potential in this gauge:
\begin{equation} \label{e44}
\widetilde V\ =\ \tanh (\gamma _1.\gamma _2 V_2)\ ,
\end{equation}
with $V_2$ given by Eq. (\ref{e43}). Expression (\ref{e44}) is
nothing but the fermionic generalization of the electromagnetic
potential proposed by Todorov in the quasipotential approach for
two spin-0 particle systems, on the basis of minimal substitution
rules \cite{t}, and later used in spectroscopic calculations
\cite{cva1,cbwva}.
It should be stressed that although potential $\widetilde V$ above is
bounded in the limit $r\rightarrow 0$, the corresponding effective
potential of the final Pauli$-$Schr\"odinger type equation of the
reduced wave function \cite{ms} has the usual $1/r$ and $1/r^2$
singularities. Furthermore, in the infinite mass limit for one of the
particles, it is only the one-photon exchange contribution that
survives in the wave equations and the latter reduce to the Dirac
equation of the finite mass particle in the presence of the Coulomb
potential.
\par
The major result of the present work is that the sum of the leading
terms of multiphoton exchange diagrams provides a local expression
for the potential in three-dimensional $x$-space. This feature
considerably simplifies the analysis and resolution of the
corresponding wave equations, without loss of the relativistic
invariance of the theory, neither of the effects of multiphoton
exchanges. These effects become more sizable in strong coupling
problems, such as evaluation of the short-distance part of the
interquark potential in QCD, which is very similar to the QED potential,
with values of the coupling constant of the order of 0.5,
or probe of the strong coupling regime in QED.
Formal generalizations to treat with the present summation method
effective interactions, in particular the confining ones, are possible
as well. The cases of spin-0 particles and of scalar interaction and a
more detailed account of this work will be presented elsewhere.
\par

\newpage

\newpage

\noindent
{\large {\bf Figures}}
\vspace{1 cm}

\noindent
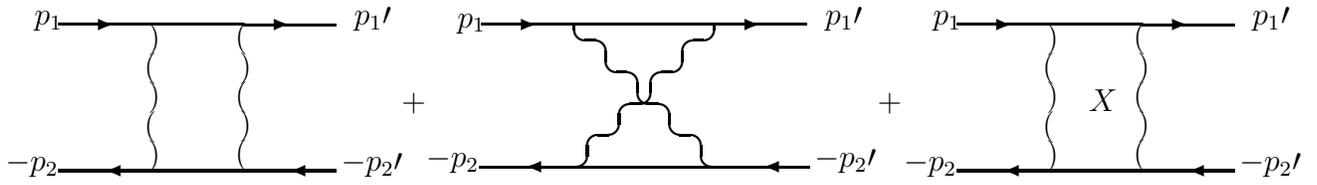
Fig. 1. Two-photon exchange diagrams; the ``constraint  diagram''
is denoted by a cross.
\vspace{0.3 cm}

\noindent
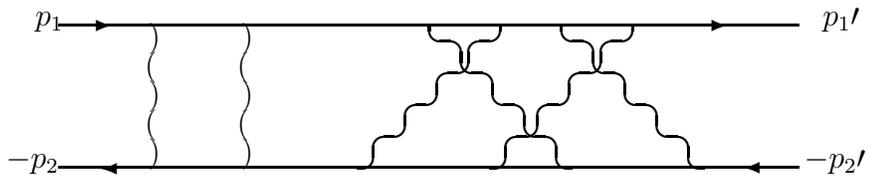
Fig. 2. A multiphoton-exchange diagram.
\vspace{0.3 cm}

\noindent
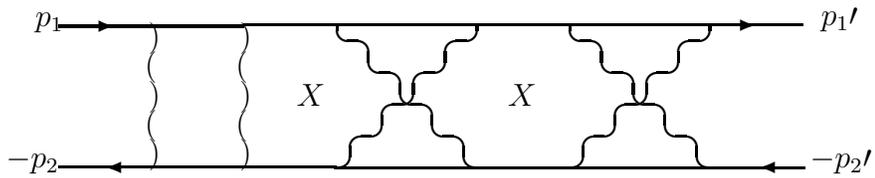
Fig. 3. A typical ``constraint diagram'' where the constraint factor
$\widetilde g_0$ [Eqs. (\protect \ref{e9}) and (\protect \ref{e11})]
appears twice ($p=2$).

\newpage

%
%
%
%
%
%

\input FEYNMAN
\renewcommand{\baselinestretch}{1.2}
\textwidth 160 mm
\textheight 240mm
\topmargin -20mm
\oddsidemargin -0.54 cm
\evensidemargin -0.54 cm

\indent

\indent

\vskip 2cm

\begin{picture}(35000,55000)
\THICKLINES
\put(19000,43000){Fig. 1}
\drawline\fermion[\E\REG](1000,53000)[3500]
\drawarrow[\LDIR\ATBASE](\pmidx,\pmidy)
\global\advance\pfrontx by -900
\put(\pfrontx,\pfronty){$p_1$}

\drawline\photon[\S\CURLY](\fermionbackx,\fermionbacky)[5]

\drawline\fermion[\E\REG](\fermionbackx,\fermionbacky)[3500]

\drawline\fermion[\W\REG](\photonbackx,\photonbacky)[3500]
\drawarrow[\W\ATTIP](\pmidx,\pmidy)
\global\advance\pbackx by -2000
\put(\pbackx,\pbacky){${-p_2}$}

\drawline\fermion[\E\REG](\photonbackx,\photonbacky)[3500]

\drawline\photon[\N\CURLY](\fermionbackx,\fermionbacky)[5]

\drawline\fermion[\E\REG](\fermionbackx,\fermionbacky)[3500]
\drawarrow[\W\ATTIP](\pmidx,\pmidy)
\global\advance\pbackx by 200
\put(\pbackx,\pbacky){${-p_2\prime}$}

\global\advance\photonbacky by 30
\drawline\fermion[\E\REG](\photonbackx,\photonbacky)[3500]
\drawarrow[\E\ATTIP](\pmidx,\pmidy)
\global\advance\pbackx by 700
\put(\pbackx,\pbacky){$p_1\prime$}

\put(14000,49700){$+$}

\drawline\fermion[\E\REG](17000,53000)[3500]
\drawarrow[\LDIR\ATBASE](\pmidx,\pmidy)
\global\advance\pfrontx by -900
\global\advance\pfronty by -100
\put(\pfrontx,\pfronty){$p_1$}

\drawline\photon[\SE\FLIPPEDFLAT](\fermionbackx,\fermionbacky)[9]

\drawline\fermion[\E\REG](\fermionbackx,\fermionbacky)[5500]

\drawline\fermion[\E\REG](\photonbackx,\photonbacky)[3500]
\drawarrow[\W\ATTIP](\pmidx,\pmidy)
\global\advance\pbackx by 200
\global\advance\pfronty by -100
\put(\pbackx,\pbacky){${-p_2\prime}$}

\drawline\fermion[\W\REG](\photonbackx,\photonbacky)[5500]

\drawline\photon[\NE\FLIPPEDFLAT](\fermionbackx,\fermionbacky)[9]

\drawline\fermion[\W\REG](\photonfrontx,\photonfronty)[3500]
\drawarrow[\W\ATTIP](\pmidx,\pmidy)
\global\advance\pbackx by -2000
\global\advance\pfronty by -100
\put(\pbackx,\pbacky){$-p_2$}

\drawline\fermion[\E\REG](\photonbackx,\photonbacky)[3500]
\drawarrow[\E\ATTIP](\pmidx,\pmidy)
\global\advance\pbackx by 700
\global\advance\pfronty by -100
\put(\pbackx,\pbacky){$p_1\prime$}


\put(32000,49700){$+$}

\put(40000,49700){$X$}
\drawline\fermion[\E\REG](35000,53000)[3500]
\drawarrow[\LDIR\ATBASE](\pmidx,\pmidy)
\global\advance\pfrontx by -900
\put(\pfrontx,\pfronty){$p_1$}

\drawline\photon[\S\CURLY](\fermionbackx,\fermionbacky)[5]

\drawline\fermion[\E\REG](\fermionbackx,\fermionbacky)[3500]

\drawline\fermion[\W\REG](\photonbackx,\photonbacky)[3500]
\drawarrow[\W\ATTIP](\pmidx,\pmidy)
\global\advance\pbackx by -2000
\put(\pbackx,\pbacky){${-p_2}$}

\drawline\fermion[\E\REG](\photonbackx,\photonbacky)[3500]

\drawline\photon[\N\CURLY](\fermionbackx,\fermionbacky)[5]

\drawline\fermion[\E\REG](\fermionbackx,\fermionbacky)[3500]
\drawarrow[\W\ATTIP](\pmidx,\pmidy)
\global\advance\pbackx by 200
\put(\pbackx,\pbacky){${-p_2\prime}$}

\global\advance\photonbacky by 30
\drawline\fermion[\E\REG](\photonbackx,\photonbacky)[3500]
\drawarrow[\E\ATTIP](\pmidx,\pmidy)
\global\advance\pbackx by 700
\put(\pbackx,\pbacky){$p_1\prime$}
\put(19000,23000){Fig. 2}
\drawline\fermion[\E\REG](7000,33000)[28000]
\global\advance\pfrontx by -900
\put(\pfrontx,\pfronty){$p_1$}
\global\advance\pbackx by 900
\put(\pbackx,\pbacky){$p_1\prime$}

\drawline\photon[\S\CURLY](10500,33090)[5]
\global\advance\photonfrontx by -1500
\drawarrow[\E\ATTIP](\photonfrontx,33000)
\global\advance\photonbackx by -1900
\global\advance\photonbacky by 50
\drawarrow[\W\ATTIP](\photonbackx,\photonbacky)

\global\advance\photonbacky by 20
\drawline\fermion[\E\REG](7000,\photonbacky)[28000]
\global\advance\pfrontx by -2000
\put(\pfrontx,\pfronty){${-p_2}$}
\global\advance\pbackx by 200
\put(\pbackx,\pbacky){$-p_2\prime$}

\drawline\photon[\S\CURLY](14000,33090)[5]
\global\advance\photonfrontx by -1500
\global\advance\photonbackx by -1500
\global\advance\photonbacky by 50

\drawline\photon[\SE\FLIPPEDFLAT](21000,33000)[9]
\global\advance\photonfrontx by -2500
\global\advance\photonbackx by -1500

\drawline\photon[\SW\FLAT](23700,33000)[9]
\global\advance\photonfrontx by -900
\global\advance\photonbackx by -2300

\drawline\photon[\SE\FLIPPEDFLAT](26000,33000)[9]
\global\advance\photonfrontx by -600
\global\advance\photonbackx by -3000
\global\advance\photonbackx by 4500
\drawarrow[\W\ATTIP](\photonbackx,\photonbacky)

\drawline\photon[\SW\FLAT](28700,33000)[9]
\global\advance\photonfrontx by -1100
\global\advance\photonfrontx by 4700
\drawarrow[\E\ATTIP](\photonfrontx,33000)
\global\advance\photonbackx by -1900
\put(19000,1000){Fig. 3}
\put(16000,8000){$X$}
\put(24000,8000){$X$}
\drawline\fermion[\E\REG](7000,11000)[3500]
\drawarrow[\LDIR\ATBASE](\pmidx,\pmidy)
\global\advance\pfrontx by -900
\put(\pfrontx,\pfronty){$p_1$}

\global\advance\fermionbacky by 110
\drawline\photon[\S\CURLY](\fermionbackx,\fermionbacky)[5]
\global\advance\photonbacky by 90

\global\advance\fermionbacky by -110
\drawline\fermion[\E\REG](\fermionbackx,\fermionbacky)[3500]

\drawline\fermion[\W\REG](\photonbackx,\photonbacky)[3500]
\drawarrow[\W\ATTIP](\pmidx,\pmidy)
\global\advance\pbackx by -2000
\put(\pbackx,\pbacky){${-p_2}$}

\drawline\fermion[\E\REG](\photonbackx,\photonbacky)[3500]

\global\advance\fermionbacky by -110
\drawline\photon[\N\FLIPPEDCURLY](\fermionbackx,\fermionbacky)[5]
\global\advance\fermionbacky by 110

\drawline\fermion[\E\REG](\fermionbackx,\fermionbacky)[3500]

\drawline\fermion[\E\REG](\photonbackx,\photonbacky)[3500]

\drawline\photon[\SE\FLIPPEDFLAT](\fermionbackx,\fermionbacky)[9]

\drawline\fermion[\E\REG](\fermionbackx,\fermionbacky)[5500]

\drawline\fermion[\E\REG](\photonbackx,\photonbacky)[3500]

\drawline\fermion[\W\REG](\photonbackx,\photonbacky)[5500]

\drawline\photon[\NE\FLIPPEDFLAT](\fermionbackx,\fermionbacky)[9]

\drawline\fermion[\E\REG](\photonbackx,\photonbacky)[3500]

\drawline\photon[\SE\FLIPPEDFLAT](\fermionbackx,\fermionbacky)[9]

\drawline\fermion[\E\REG](\fermionbackx,\fermionbacky)[5500]

\drawline\fermion[\E\REG](\photonbackx,\photonbacky)[3500]
\drawarrow[\W\ATTIP](\pmidx,\pmidy)
\global\advance\pbackx by 200
\put(\pbackx,\pbacky){$-p_2\prime$}

\drawline\fermion[\W\REG](\photonbackx,\photonbacky)[5500]

\drawline\photon[\NE\FLIPPEDFLAT](\fermionbackx,\fermionbacky)[9]

\drawline\fermion[\E\REG](\photonbackx,\photonbacky)[3500]
\drawarrow[\E\ATTIP](\pmidx,\pmidy)
\global\advance\pbackx by 700
\put(\pbackx,\pbacky){$p_1\prime$}

\end{picture}


\begin{thebibliography}{50}

\bibitem{ll}G. Longhi and L. Lusanna, eds., {\it Constraint's
Theory and Relativistic Dynamics}, proceedings of the Firenze Workshop,
1986 (World Scientific, Singapore, 1987).
\bibitem{cva1}H.W. Crater and P. Van Alstine, Phys. Rev. D 36 (1987)
3007; {\it ibid.} 37 (1988) 1982; J. Math. Phys. 31 (1990) 1998.
\bibitem{s1}H. Sazdjian, Phys. Rev. D 33 (1986) 3401; J. Math. Phys.
29 (1988) 1620.
\bibitem{by}G.T. Bodwin and D.R. Yennie, Phys. Rep. C 43 (1978) 267;\\
G.P. Lepage, Phys. Rev. A 16 (1977) 863;\\
R. Barbieri and E. Remiddi, Nucl. Phys. B141 (1978) 413.
\bibitem{l}S. Love, Ann. Phys. (N.Y.) 113 (1978) 153.
\bibitem{bcm}R. Barbieri, M. Ciafaloni and P. Menotti, Nuovo Cimento 55 A
(1968) 701.
\bibitem{t}I.T. Todorov, Phys. Rev. D 3 (1971) 2351; in {\it Properties
of Fundamental Interactions}, ed. A. Zichichi (Editrice Compositori,
Bologna, 1973) Vol. 9, Part C, p. 931.
\bibitem{rta}V.A. Rizov, I.T. Todorov and B.L. Aneva, Nucl. Phys. B98
(1975) 447.
\bibitem{ms}J. Mourad and H. Sazdjian, J. Math. Phys. 35 (1994) 6379.
\bibitem{s2}H. Sazdjian, in {\it
Extended Objects and Bound Systems}, proceedings of the Karuizawa
International Symposium, 1992, eds. O. Hara, S. Ishida and S. Naka
(World Scientific, Singapore, 1992), p. 117.
\bibitem{cw}H. Cheng and T.T. Wu, Phys. Rev. 182 (1969) 1852; {\it ibid.}
186 (1969) 1611.
\bibitem{ls}M. L\'evy and J. Sucher, Phys. Rev. 186 (1969) 1656.
\bibitem{b}S.J. Brodsky, in Brandeis Lectures 1969, eds. M. Chr\'etien
and E. Lipworth (Gordon and Breach, New York, 1971), p. 93.
\bibitem{be} H. Bateman and A. Erd\'elyi, {\it Higher
Transcendental Functions} (McGraw$-$Hill, New York, 1953), Vol. 1,
p. 104.
\bibitem{cbwva}H.W. Crater, R.L. Becker, C.Y. Wong and P. Van Alstine,
Phys. Rev. D 46 (1992) 5117.

\end{thebibliography}
\end{document}